\documentclass[12pt]{article}
\usepackage{graphicx}

\newcommand{\beq}{\begin{equation}}
\newcommand{\eeq}{\end{equation}}
\newcommand{\beqn}{\begin{eqnarray}}
\newcommand{\eeqn}{\end{eqnarray}}
\newcommand{\dddot}[1]{#1\hspace{-0.5em}\raisebox{0.6em}{$.\hspace{-0.07em}.\hspace{-0.07em}.$}}

\newcommand{\br}{\mbox{${\mathbf r}$}}

\newcommand{\De}{\mbox{${\Delta}$}}

\newcommand{\ep}{\mbox{${\varepsilon}$}}
\newcommand{\om}{\mbox{${\omega}$}}

\begin{document}
\begin{center}

{\bf \large Approximate Selection Rule for Orbital Angular
Momentum\\ in Atomic Radiative Transitions}

\vspace{0.5cm}

I.B. Khriplovich and D.V. Matvienko

\vspace{0.5cm}

{\it Budker Institute of Nuclear Physics,\\} {\it 630090
Novosibirsk, Russia,\\} {\it and Novosibirsk University }
\end{center}

\begin{abstract}
We demonstrate that radiative transitions with $\De l = - 1$ are
strongly dominating for all values of $n$ and $l$, except small
region where $l \ll n$.
\end{abstract}

It is well-known that the selection rule for the orbital angular
momentum $l$ in electromagnetic dipole transitions, dominating in
atoms, is $\De l = \pm 1$, i. e. in these transitions the angular
momentum can both increase and decrease by unity. Meanwhile, the
classical radiation of a charge in the Coulomb field is always
accompanied by the loss of angular momentum. Thus, at least in the
semiclassical limit, the probability of dipole transitions with
$\De~l~=~-~1$ is higher. Here we discuss the question how strongly
and under what exactly conditions the transitions with $\De l = -
1$ dominate in atoms. (To simplify the presentation, we mean
always, here and below, the radiation of a photon, i. e.
transitions with $\De n < 0$. Obviously, in the case of photon
absorption, i. e. for $\De n > 0$, the angular momentum
predominantly increases.)

The analysis of numerical values for the transition probabilities
in hydrogen presented in \cite{bs} has demonstrated that even for
$n$ and $l$, comparable with unity, i. e. in a nonclassical
situation, radiation with $\De l = - 1$ can be much more probable
than that with $\De l = 1$.

Later, the relation between the probabilities of transitions with
$\De l = - 1$ and $\De l = 1$ was investigated in \cite{dk} by
analyzing the corresponding matrix elements in the semiclassical
approximation. The conclusion made therein is also that the
transitions with $\De l = - 1$ dominate, and the dominance is
especially strong when $l > n^{2/3}$.

Here we present a simple solution of the problem using the
classical electrodynamics and, of course, the correspondence
principle. Our results describe the situation not only in the
semiclassical situation. Remarkably enough, they agree, at least
qualitatively,  with the results of \cite{bs}, although the latter
refer to transitions with $|\De n| \sim n \sim 1$ and $l \sim 1$,
which are not classical at all.

We start our analysis with a purely classical problem. Let a
particle with a mass $m$ and charge $\;-\,e$ moves in an
attractive Coulomb field, created by a charge $e$, along an
ellipse with large semi-axis $a$ and eccentricity $\ep$. It is
known \cite{ll2} that the radiation intensity at a given harmonic
$\nu$ is here
\beq
I_{\nu} = \,\frac{4e^2\om_0^4 \nu^4 a^2}{3c^3}\left(\xi_{\nu}^2 +
\eta_{\nu}^2\right);
\eeq
\beq\label{xet}
\xi_{\nu} = \,\frac{1}{\nu}\,J_{\nu}^{\prime}(\nu\ep), \quad
\eta_{\nu}= \,\frac{\sqrt{1-\ep^2}}{\nu\ep}\,J_{\nu}(\nu\ep).
\eeq
In expressions (\ref{xet}), $J_{\nu}(\nu\ep)$ is the Bessel
function, and $J_{\nu}^{\prime}(\nu\ep)$ is its derivative. We use
the Fourier transformation in the following form:
\[
x(t) = a\sum^{\infty}_{\nu=-\infty} \xi_{\nu}\,e^{i\nu\omega_0 t}
= 2a\sum^{\infty}_{\nu =0} \xi_{\nu} \cos \nu\om_0 t,
\]
\[
y(t) = a\sum^{\infty}_{\nu=-\infty} \eta_{\nu}\,e^{i\nu\omega_0 t}
= 2a\sum^{\infty}_{\nu =0} \eta_{\nu} \sin \nu\om_0 t,
\]
where all dimensionless Fourier components $\xi_{\nu}$ and
$\eta_{\nu}$ are real, and $\xi_{-\nu} = \xi_{\nu}$,
$\eta_{-\nu}~=~-~\eta_{\nu}$. We note that the Cartesian
coordinates $x$ and $y$ are related here to the polar coordinates
$r$ and $\phi$ as follows: $x=r\cos\phi, \; y=r\sin\phi$, where
$\phi$ increases with time. Thus, the angular momentum is directed
along the $z$ axis (but not in the opposite direction).

We note also that, since $0~\leq~\ep~\leq~1$, both
$J_{\nu}(\nu\ep)$ and $J_{\nu}^{\prime}(\nu\ep)$ are reasonably
well approximated by the first term of their series expansion in
the argument. Therefore, all the Fourier components $\xi_{\nu}$
and $\eta_{\nu}$ are positive.

In the quantum problem (where $\nu = |\De n|$), the probability of
transition in the unit time is
\beq\label{wnu}
W_{\nu} = \frac{I_{\nu}}{\hbar \om_0\nu} =\,\frac{4e^2\om_0^3
\nu^3 a^2}{3c^3\hbar}\left(\xi_{\nu}^2 + \eta_{\nu}^2\right),
\quad \om_0 = \frac{me^4}{\hbar^2 n^3}\,.
\eeq

Now, the loss of angular momentum with radiation is \cite{ll2}
\[
\dot{\bf M}=\,\frac{2e^2}{3c^3}\,\br \times \bf{\dddot{r}}\,.
\]
Going over here to the Fourier components, we obtain
\[
\dot{\bf M}_{\nu}=\,-\,\frac{4 e^2 \om_0^2 \nu^2}{3c^3}\,\br_{\nu}
\times \dot{\bf r}_{\nu}\,,
\]
or (with our choice of the direction of coordinate axes, and with
the angular momentum measured in the units of $\hbar$)
\beq\label{dmnu}
\dot{M}_{\nu}\,=\,-\,\frac{4e^2 \om_0^3 \nu^3
a^2}{3c^3\hbar}\,2\xi_{\nu}\eta_{\nu}\,.
\eeq

Obviously, the last expression is nothing but the difference
between the probabilities of transitions with $\De l = 1$ and $\De
l = - 1$ in the unit time:
\beq\label{w-}
\dot{M}_{\nu}\,= W_{\nu}^+ - W_{\nu}^-.
\eeq
Of course, the total probability (\ref{wnu}) can be written as
\beq\label{w+}
W_{\nu} = W_{\nu}^+ + W_{\nu}^-.
\eeq
From explicit expressions (\ref{wnu}) and (\ref{dmnu}) it is clear
that inequality $W_{\nu}^+ \ll W_{\nu}^-$ holds if
$2\xi_{\nu}\eta_{\nu} \approx \xi^2_{\nu} + \eta^2_{\nu}$, or
$\eta_{\nu} \approx \xi_{\nu}$. The last relation is valid for
$\ep \ll 1$, i. e. for orbits close to circular ones. (The
simplest way to check it, is to use in formulae (\ref{xet}) the
explicit expression for the Bessel function at small argument:
$J_{\nu}(\nu\ep)=\,(\nu\ep)^\nu/(2^\nu\nu\,!)$.)

This conclusion looks quite natural from the quantum point of
view. Indeed, it is the state with the orbital quantum number $l$
equal to $n-1$ (i. e. with the maximum possible value for given
$n$) which corresponds to the circular orbit. In result of
radiation $n$ decreases, and therefore $l$ should decrease as
well.

The surprising fact is, however, that in fact the probabilities
$W_{\nu}^-$ of transitions with $\De l = -1$ dominate numerically
everywhere, except small vicinity of the maximum possible
eccentricity $\ep = 1$. For instance, if $\ep \simeq 0.9$ (which
is much more close to 1 than to~0\,!), then at~$\nu~=~1$ the
discussed probability ratio is very large, it constitutes
\[
\frac{W_{\nu}^-}{W_{\nu}^+}\,\simeq 12\,.
\]
The change with $\ep$ of the ratio of $W_{\nu}^+$ to $W_{\nu}^-$
for two values of $\nu$ is illustrated in Fig.~1.
\begin{figure}[h]
\centering
\includegraphics[width=0.8\linewidth]{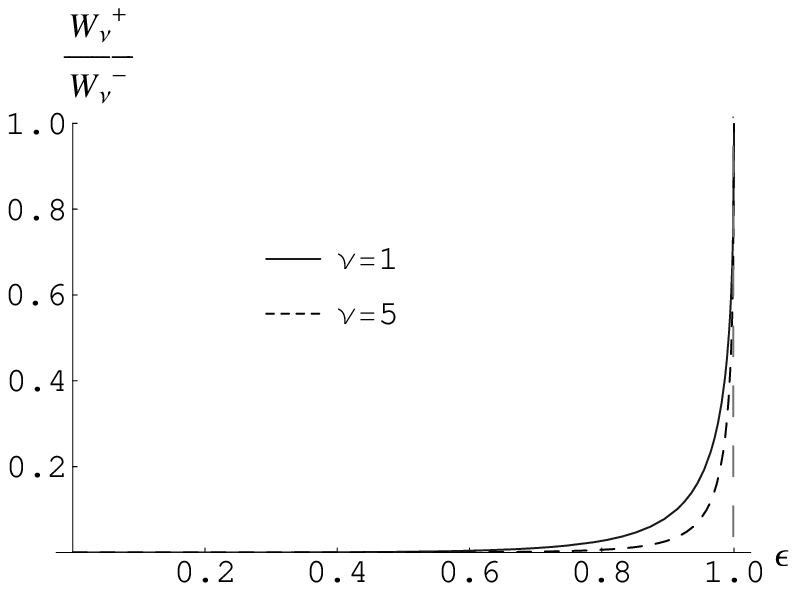}
\begin{center} {\em Fig. 1} $\;\;\;\;\;\;\;\;\;\;\;\;\;\;\;$ \end{center}
\end{figure}
The curves therein demonstrate in particular that with the
increase of $\nu$, the region where $W_{\nu}^-$ and $W_{\nu}^+$
are comparable, gets more and more narrow, i. e. when $\nu$ grows,
the corresponding curves tend more and more to a right angle.

Let us go over now to the quantum problem. In the semiclassical
limit, the classical expression for the eccentricity
\beq\label{ex}
\ep =\sqrt{1 +\,\frac{2EM^2}{me^4}}
\eeq
is rewritten with usual relations $E=-me^4/(2\hbar^2 n^2)$ and
$M=\hbar l$ as
\beq\label{semi}
\ep =\sqrt{1 -\,\frac{l^2}{n^2}}\,.
\eeq
In fact, the exact expression for $\ep$, valid for arbitrary $l$
and $n$, is \cite{ll2}:
\beq\label{semi1}
\ep =\sqrt{1 -\,\frac{l(l+1)+1}{n^2}}\,.
\eeq
Clearly, in the semiclassical approximation the eccentricity is
close to unity only under condition $l \ll n$. If this condition
does not hold, one may expect that in the semiclassical limit the
transitions with $\De l = -1$ dominate. In other words, as long as
$l \ll n$, the probabilities of transitions with decrease and
increase of the angular momentum are comparable. But if the
angular momentum is not small, it is being lost predominantly in
radiation. This situation looks quite natural.

The next point is that with the increase of $|\De n| = \nu$, the
region where $W_{\nu}^-$ and $W_{\nu}^+$ are comparable, gets more
and more narrow in agreement with the observation made in
\cite{dk}.

However, we do not see any hint at some special role (advocated in
\cite{dk}) of the condition $l > n^{2/3}$ for the dominance of
transitions with $\De l = - 1$.

As mentioned already, the analysis of the numerical values of
transition probabilities~\cite{bs} demonstrates that even for $n$
and $l$ comparable with unity and $|\De n|\simeq n$, i. e. in the
absolutely nonclassical regime, the transitions with $\De l = - 1$
are still much more probable than those with $\De l = 1$. The
results of this analysis for the ratio $W^-/W^+$ in some
\begin{table}[h]
\begin{center}
\begin{tabular}{|c|c|c|c|c|c|c|c|c|} \hline
 & & & && & \\
&$\frac{W_{4p \to 3s}}{W_{4p \to 3d}}$ &$\frac{W_{5p \to
4s}}{W_{5p \to 4d}}$&$\frac{W_{5d \to 4p}}{W_{5d \to
4f}}$&$\frac{W_{6f \to 5d}}{W_{6f \to 5g}}$&$\frac{W_{5p \to
3s}}{W_{5p\to 3d}}$&$\frac{W_{6p \to 3s}}{W_{6p\to 3d}}$
 \\
 & & & && & \\ \hline
 & & & && & \\
exact  & & & && & \\
value  &10  &3.75  &28 &72 &10.67& 13.7\\
& & & && & \\ \hline
& & &  && & \\
$\bar{\ep}$& 0.87  & 0.92 & 0.81 & 0.75 & 0.90 & 0.92\\
& & & && &\\ \hline
& & &  && &\\
$\nu = |\De n|$ & 1& 1 &1  &1 &2 &3\\
 & & & && &\\ \hline
& & &  && &\\
semiclassical & & &  && &\\
value  & 17.6 & 8.7 & 34 & 58 & 17.2& 15.7\\
 & & & && &\\ \hline
\end{tabular}

\vspace{5mm} Table 3.1
\end{center}
\end{table}
transitions are presented in Table 3.1 (first line). Then we
indicate in Table 3.1 (last line) the values of these ratios
obtained in the na\"{\i}ve (semi)classical approximation. Here for
the eccentricity $\bar{\ep}$ we use the value of expression
(\ref{semi1}), calculated with $l$ corresponding to the initial
state; as to $n$, we take its value average for the initial and
final states.

The table starts with the smallest possible quantum numbers where
the transitions, which differ by the sign of $\De l$, occur, i. e.
with the ratio $W_{4p \to 3s}/W_{4p \to 3d}$. This table
demonstrates that the ratio of the classical results to the exact
quantum-mechanical ones remains everywhere within a factor of
about two. In fact, if one uses as $\bar{\ep}$ expression
(\ref{semi}), calculated in the analogous way, the numbers in the
last line change considerably. It is clear, however, that the
classical approximation describes here, at least qualitatively,
the real situation.

\end{document}